\documentclass[9pt,twocolumn,twoside]{osajnl}

\journal{ao} 

\setboolean{shortarticle}{false}

\usepackage{lineno}

\begin{document}

\title{Ultrafast Structured Light through Nonlinear Frequency Generation in an Optical Enhancement Cavity}

\author[1]{Walker M. Jones}
\author[1,*]{Melanie A. R. Reber}

\affil[1]{Department of Chemistry, University of Georgia, Athens, GA 30606}

\affil[*]{Corresponding author: mreber@uga.edu}

\begin{abstract}

The generation of shaped laser beams, or structured light, is of interest in a wide range of fields, from microscopy to fundamental physics. There are several ways to make shaped beams, most commonly using spatial light modulators comprised of pixels of liquid crystals. These methods have limitations on the wavelength, pulse duration, and average power that can be used. Here we present a method to generate shaped light that can be used at any wavelength from the UV to IR, on ultrafast pulses, and a large range of optical powers. By exploiting the frequency difference between higher order modes, a result of the Gouy phase, and cavity mode matching, we can selectively couple into a variety of pure and composite higher order modes. Optical cavities are used as a spatial filter and then combined with sum frequency generation in a nonlinear crystal as the output coupler to the cavity to create ultrafast, frequency comb structured light. 

\end{abstract}

\setboolean{displaycopyright}{false}

\maketitle

\begin{center}\date{May 20, 2024}\end{center}

Structured light has applications in many areas of science, including microscopy\cite{Zeng_2023}, quantum information\cite{Kagalwala_NatComm2017}, and fundamental physics\cite{Forbes_AdvOptPhoton2016}. Shaped optical beams have an additional degree of freedom, the spatial mode, and can be used for things like multimodal entanglement and to decrease the quantum noise\cite{Wagner_Science2008,Lassen_PRL2007}. As such, there are a wide range of ways to make shaped beams including spatial light modulators\cite{Li_Science2019}, q-plates\cite{Rubano_JOSAB2019}, and even in the laser itself\cite{Forbes_NatPhoton2021,Forbes_review}. Spatial modification of optical beams is most commonly achieved with spatial light modulators where an array of pixels covers the extent of the beam and the shape is modified by turning the pixels on and off. These devices are limited in the resolution of the pixels, which are typically made up of liquid crystals\cite{Li_Science2019} with liquid crystals on silicon as the most common and commercially available\cite{Lazarev_OE2019}. There are also power and wavelength limitations to these methods. The other broad category of methods to create structured light is to generate it directly in the laser cavity. These methods often place an obstruction in the laser cavity to force the cavity to lase in a non-$TEM_{00}$ mode. However, achieving the desired spatial mode while working with the complex nonlinear gain dynamics adds a level of complexity to these methods, and the achievable wavelengths are limited by the gain medium.

Here we present a method to generate shaped light that can be used at any wavelength from the UV to IR, on ultrafast pulses, at a large range of optical powers. Optical cavities are used as the spatial filter and generate shaped laser beams through sum frequency generation in a nonlinear crystal as the output coupler to the cavity. By exploiting the frequency difference between higher order modes, a result of the Gouy phase, and cavity mode matching we selectively couple into a variety of pure and composite higher order modes. Since the cavities are not part of the laser cavity, there is no concern about any effects of or dynamics from the gain medium.

The use of a cavity as a spatial filter and the impact of the cavity on an image was first explored theoretically and experimentally by Gigan at al\cite{Gigan_PRA2005}. They input an image to the cavity, including a slit and series of slits, and looked at the resultant image after passing through their hemiconfocal cavity. They showed the cavity-transmitted beam was a self-Fourier transform of the image. In this letter, we use the cavity modes as a spatial filter to shape the incoming light. It has been known for many years that optical cavities support higher order Hermite-Gaussian beams,\cite{Kogelnik_ApplOpt1966} and that those modes come on resonance with the cavity at different frequencies because of the Gouy phase. Higher-order modes were generally avoided and it is only relatively recently that these have started to be used and researched\cite{Weitenberg_OE2011,Djevahirdjian_IOP2020}. Using higher order cavity modes, brought on by the placement of a wire obstruction in the cavity, Weitenberg et al characterized the field with the obstruction in place with the goal to explore the use of a mirror with a hole in it for output of high harmonic generated light\cite{Weitenberg_JoO2015,Weitenberg_OE2011}. Higher order modes of optical cavities have also been used to generate squeezed states\cite{Lassen_PRL2007}. 

In this work, we will exploit the higher order modes to shape the incoming beam. It is necessary to use a frequency comb laser to coupled ultrafast light into an optical cavity\cite{Reber_optica2016}. To get the shaped light out of the cavity, we use nonlinear frequency generation in a beta-barium borate (BBO) crystal. Nonlinear frequency generation between frequency comb lasers produces an optical frequency comb, which is well documented\cite{Schliesser_NatPhoton2012}. While nonlinear frequency generation has been used in a range of applications for many years, however there are only a few detailed investigations of nonlinear frequency generation with higher order modes\cite{Buchhave_OE2008,Delaubert_OE2007,Qi_AppPhysB2022}. Using non-linear frequency generation to output the modes from the cavity has not been explored before, to the best of our knowledge.

\begin{figure}[ht]
\centering\includegraphics [width=0.45\textwidth]{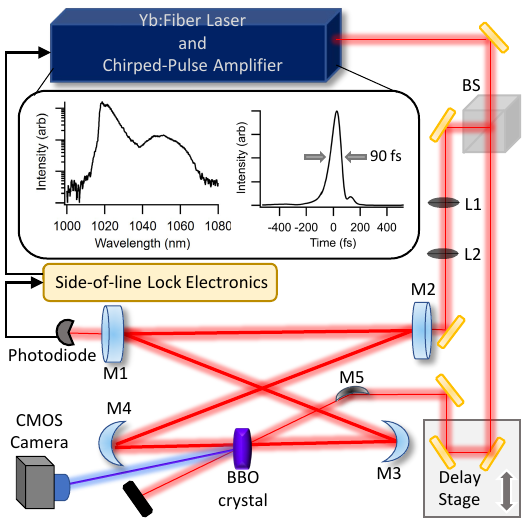}
\caption{Experimental setup and laser characterization. A homebuilt Ytterbium:fiber laser and chirped-pulse amplifier supply 90 fs pulses at 75 MHz repetition rate centered at 1040 nm. The inset shows an optical spectra and pulse reconstruction. The beam is split into two, one is locked to a bowtie enhancement cavity and the other sent to a delay stage. A beta-barium borate (BBO) crystal is placed at the main focus of the cavity where the non-cavity beam, fused by a curved mirror M5, intersects. The resultant SHG beam is sent to a CMOS camera and photodiode for detection.}
\label{layout}
\end{figure}

In this work, a homebuilt ytterbium fiber (Yb:fiber) laser frequency comb and amplification system generates 250 mW centered at 1040 nm, 75 MHz repetition rate, and 100 fs pulses. The Yb:fiber laser and amplifier are described in details elsewhere\cite{Cooper_ApplOpt,Walker_inprep} and will only briefly be described here. The Yb:fiber oscillator produces about 40 mW pulses of a few picoseconds duration at 75 MHz repetition rate. The output of the laser is coupled into SM980 fiber, which is then spliced to 15 cm of a core pumped, single clad Yb:doped fiber (Thorlabs YB1200-4/125). The doped fiber is forward-pumped by a 500 mW 970 nm diode laser and the amplified output is then compressed in a double-pass, transmission grating compressor to 100 fs with 250 mW average power. The amplified output is then split into two paths. One path goes through a set of two lenses to be mode-matched to the optical cavity. The cavity is in a bowtie configuration consisting of two flat mirrors and two curved mirrors with 60 cm focal lengths. The length of the external cavity matches the repetition rate of the laser. The second beam path travels through a computer-controlled delay stage and 25cm focal length curved mirror. The intra- and extra-cavity focii are spatially overlapped by aligning both through a 25$\mu$m pinhole. 

The cavity is locked to the frequency comb using a side-of-line lock. The light transmitted from one of the flat cavity mirrors is incident on a photodetector. The photodetector signal goes to a loop filter board, where it is compared to an adjustable DC voltage. The DC voltage is chosen such that it intersects the transmission peak about two-thirds the way up the peak. This type of lock was chosen over a Pound-Drever-Hall lock or top-of-line lock because of the ease of locking to different modes with large differences in transmitted light, however a Pound-Drever-Hall lock can also be used. To lock to a different $TEM$ cavity mode involves tuning the length of the cavity to correspond to the desired $TEM$ mode frequency and then activating the lock.

At the focus of the cavity between the two curved mirrors is a 0.5 mm Beta-Barium Borate (BBO) crystal. The second beam crosses the cavity beam at the BBO crystal in an auto-correlation geometry. The two beams undergo second harmonic generation in the crystal and a spatially separated second harmonic beam is emitted in the $\Vec{k}_{intra} + \Vec{k}_{extra}$ direction. The generated second harmonic beam is spatially separated from the two infrared beams and sent to a CMOS camera and an amplified photodiode for detection. 

\begin{figure}[ht]
\centering\includegraphics [width=0.43\textwidth]{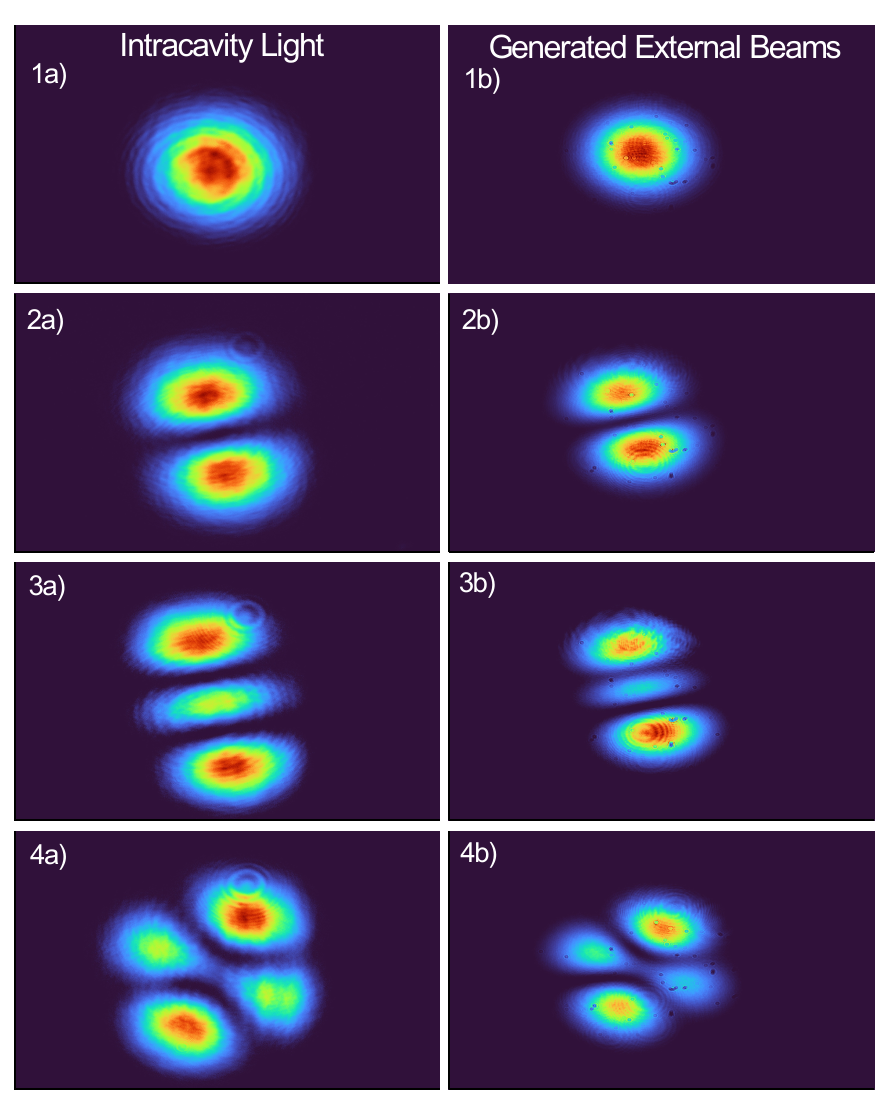}
\caption{Pictures of the intracavity modes (a) and the corresponding mode of the generated SHG beam (b). All of these were taken with vertical misaligment into the cavity and a probe beam diameter of 125 $\mu m$, which is larger than the cavity beam when they cross. Nominal mode assignments based upon beam shape are as follows: Panel 1 is the $TEM_{00}$ beam, 2 is a $TEM_{10}$ beam, 3 is a $TEM_{20}$ beam, and 4 is $TEM_{11}$ beam.}
\label{modephotos}
\end{figure}

\begin{figure}[ht!]
\centering\includegraphics [width=0.45\textwidth]{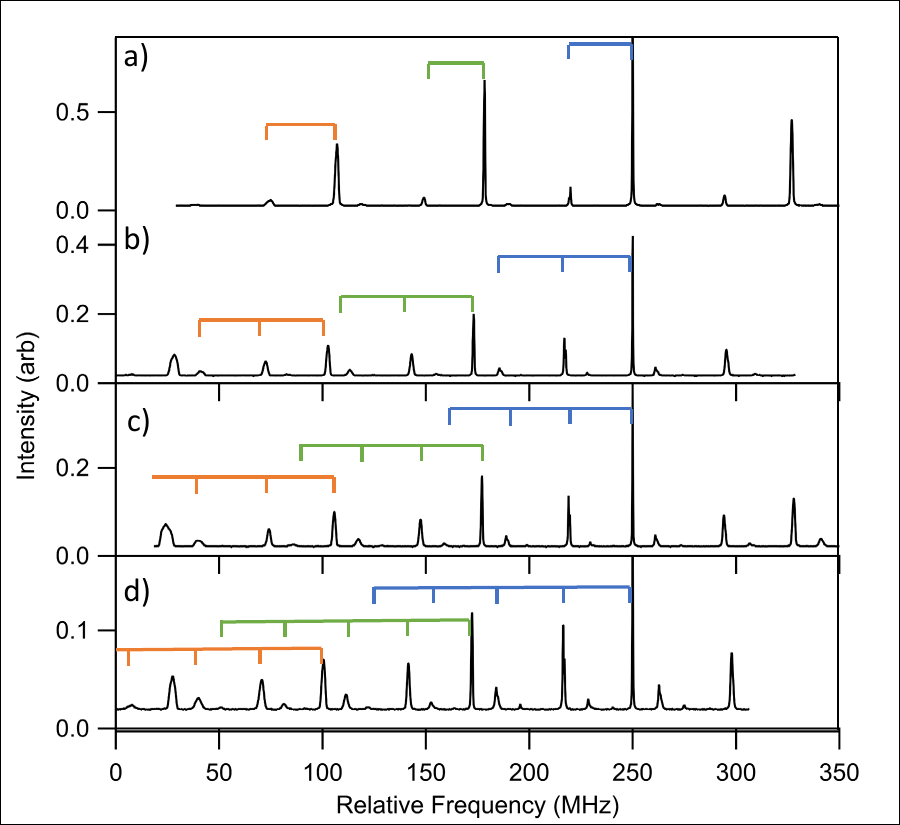}
\caption{Transmission spectra of the cavity obtained by scanning the cavity-mirror piezo. The ramp travels through multiple laser repetition rates, with the strongest labeled with the blue track, followed by a green track and then an orange track. Panel a) shows a spectra of the cavity with very good aligment, so only one higher order mode appears for each repetition rate. Panels b), c), and d) show the cavity spectra as the input beam is increasingly vertically misaligned causing more higher order modes to appear.}
\label{samplemodes}
\end{figure}

By slightly misaligning the input beam of the cavity and varying the distance between the mode-matching lens, a wide range of different spatial modes become resonant in the cavity at different cavity lengths. Figure \ref{modephotos} shows a representative set of cavity modes and the corresponding generated beam mode. More cavity output beam shapes are in Supplemental Information. Vertical misalignment yields more rectangular modes while coupling a larger beam leads to more higher order modes modes with circular symmetry. By scanning the cavity length using a piezo-driven translation stage, we can effectively take the spectra of the cavity. Figure \ref{samplemodes} shows the optical spectrum of the enhancement cavity while misaligning the input cavity increasingly misaligned vertically. Panel a shows a mostly well-aligned and mode-matched cavity, you can see the repetition rate of the cavity, which is labeled in the figure. Since the x-axis is derived from the piezo voltage, it is calibrated by taking the difference between the peaks and setting that to the repetition rate of the laser as measured on the RF spectrum analyzer. The input frequency comb optical spectrum is broader than what the cavity can support, due to dispersion in air, so we don't see coupling into a single cavity mode but also several nearby cavity modes as seen in the figure \ref{samplemodes}. The piezo does not scan perfectly linearly, as can be seen by the slight variation in peak spacing between scans and the broadening of the peaks towards both ends of each scan. The scans were aligned by setting the highest peak to 250 MHz and the average ramp voltage to frequency value as the step to generate the x-axis.

Increasing the input beam mis-alignment to the cavity makes it possible to assign the peaks in the cavity spectrum of the most misaligned spectrum, Figure \ref{samplemodes} panel d). When the alignment is done in this way, the strongest modes are clear Hermite-Gaussian modes for which the frequencies can be calculated. The difference in cavity resonant frequency between higher order Hermite Gaussian modes is, where $g_1$ and $g_2$ depend upon the cavity parameters, defined as $g_n = 1 - \frac{d_n}{R}$.\cite{Nagourney}. In the bowtie geometry, R is the radius of curvature of the curved mirrors, $d_2$ is the length between the curved mirrors, and $d_1$ is the remaining length of the cavity. The difference in resonant frequency between adjacent Hermite-Gaussian modes, $\Delta n = 1$, is
\begin{equation}
 \nu_{(n+1)mq}-\nu_{mq} = \frac{cos^{-1}[\sqrt{g_1g_2}]}{\pi}\frac{c}{L}   
\end{equation} 
where L is the length of the cavity, c is the speed of light, and $g_1$ and $g_2$ are cavity parameters. It is worth noting that because the resonant frequencies of the Hermite-Gaussian modes go as $n + m + 1$ each mode with the same $n +m$ value will be degenerate. We are often able to couple predominantly into one mode based upon the incoming beam size and spatial overlap. The difference in resonant frequencies of the different modes is a result of the Gouy Phase shift. 

The cavity used here is a bowtie cavity with 61.1 $cm$ between the curved mirrors ($d_1$) and 335.5 $cm$ as the remaining cavity length ($d_2$). The curved mirrors have a 60 $cm$ focal length and all angles are around $5^{\circ}$ or less. Using the assumption the angles are small and the effects can be neglected, the $g$ parameters are  Using these cavity parameters, we calculate that adjacent higher order modes are spaced apart by 30.9 MHz. Taking five sets of cavity ramps, fitting the peaks to a Lorentzian function to find the center frequency, and averaging the center frequency, the experimentally measured spacing between modes is 30.8 (1.1) MHz with the standard deviation in parenthesis. For reference the average repetition rate measured this was was 74.3 (3.2) MHz in comparison to 75.367 MHz from the RF spectrum analyzer. This shows that the cavity is behaving as expected and the modes are assignable with the combination of frequency and beam profile. 

\begin{figure}[ht!]
\centering\includegraphics [width=0.45\textwidth]{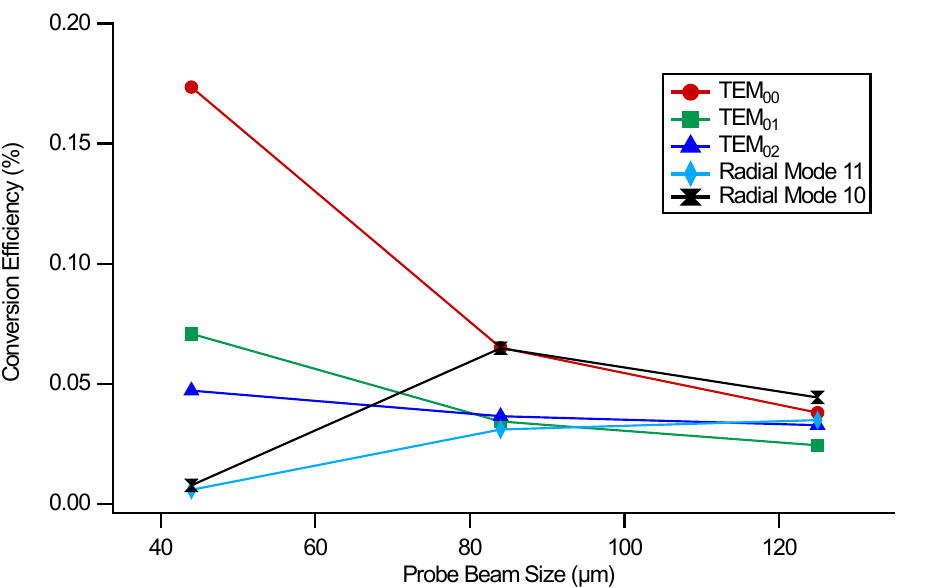}
\caption{Conversion efficiency of the different modes while locked at 2/3 peak power as a function of probe beam ($TEM_{00}$) size. The better overlap between the cavity shape and probe beam size, the higher the conversion efficiency.}
\label{efficiency}
\end{figure}

The second harmonic beam shape is a function of the overlap of the two infrared incident beams. The mode structure follows the cavity mode for the larger probe beams. As the probe beam gets smaller, the generated beam changes shape slightly to reflect the overlap with a smaller Gaussian beam. When the probe beam is large, the generated beam reproduces the cavity beam exactly, since the readout beam overfills the Gaussian beam structure, and those are the modes shown in Figure \ref{modephotos}. The efficiency of the generated light increases as the overlap increases and decreases with poorer overlap, as expected. Figure \ref{efficiency} plots the conversion efficiency as a function of probe beam size. When the probe beam is smaller than the pump beam, there is a strong dependency on beam size depending upon the cavity mode shape.

This work demonstrates the use of an optical cavity as a spatial mode filter and how it can be used to created clean, structured light. The output beam will be a frequency comb just like the two input beams\cite{Schliesser_NatPhoton2012} and have an ultrafast pulse duration\cite{Reber_optica2016}. The cavity in this work is in air, which limits the achievable pulse duration in the cavity, but could be implemented in vacuum or lower dispersion environment. This method is a versatile way to generate ultrafast shaped light. It is possible to couple light into complex mixed and higher order modes into the cavity and make complicated shapes, which could be used to generate ultrafast beams with arbitrary shapes. For example, one type of optical beam that could be generated this way is an ultrafast optical vortex beam, beams with optical angular momentum. There is increased interest in solving the problem of making ultrafast beams with orbital angular momentum (OAM)\cite{Guer_OL2024} and we plan on future extensions of the cavity technique to ultrafast and frequency comb OAM beams. There are a range of fundamental optics and physics questions that structured light could address\cite{Rubinsztein-Dunlop_2017} and further development of structured light sources, including full control of all parameters of light by taking advantage of the frequency comb nature of this technique, are planned for the future. One could also image combining two cavities, each locked to different higher-order-mode, that cross at the BBO crystal to achieve even more complicated beam shapes and structures\cite{Shen_IEEEPhoton2022}. This cavity-based method shows promise as a general method for ultrafast, structured light and structured frequency combs.

\begin{backmatter}
\bmsection{Funding} This material is based upon work supported by the National Science Foundation under Grant No. 2207784.


\bmsection{Disclosures} The authors declare no conflicts of interest.

\bmsection{Data availability} Data underlying the results presented in this paper are not publicly available at this time but may be obtained from the authors upon reasonable request.

\bmsection{Supplemental document}
See Supplement 1 for supporting content. 

\end{backmatter}

\bibliography{Arxiv_StructuredLightPaper}

\bibliographyfullrefs{Arxiv_Structured}

\end{document}